\documentclass[aip,apl,reprint,showpacs,amsmath,amssymb]{revtex4-1}

\usepackage{amsmath}

\usepackage{graphicx}

\begin{document}

\title{Quantum Beats of a Multiexciton State in Rubrene Single Crystals}

\author{Eric A. Wolf}
\author{Drew M. Finton}
\author{Vincent Zoutenbier}
\author {Ivan Biaggio}
\affiliation {Department of Physics, Lehigh University, Bethlehem, PA 18015, USA}
\date{\today}

\begin{abstract}
We observe quantum beats in the nanosecond-scale photoluminescence decay of rubrene single crystals after photoexcitation with short laser pulses in a magnetic field of 0.1 to 0.3 T. The relative amplitude of the quantum beats  is of the order of 5\%. Their  frequency is $1.3$ GHz when the magnetic field is oriented parallel to the two-fold rotation axis of the rubrene molecules and decreases to  $0.6$ GHz when the magnetic field is rotated to the crystal's molecular stacking direction.
The amplitude of the quantum beats decays alongside the non-oscillatory photoluminescence background, which at low excitation densities has an exponential decay time of $ 4.0 \pm 0.2$~ns. We interpret this as the effective lifetime of a multiexciton state  that originates from singlet-fission and can undergo geminate recombination back to the singlet state.
\end{abstract}
\maketitle

In larger organic molecules, the first electronic excited state can have a significantly lower energy in a triplet configuration when compared to the photoexcited singlet configuration. As a consequence, a singlet exciton in an organic molecular crystal can efficiently transition to a pair of  triplet excitons with total spin of zero, which can ultimately dissociate into independent, long-lived triplet excitons. The early observation of a delayed photoluminescence (PL) after photoexcitation with laser pulses first indicated that independent triplet excitons could meet and undergo a \emph{fusion} process to produce a radiative singlet exciton\cite{Kepler63, Hall63}, and it was soon recognized that a \emph{fission} process in which a singlet exciton decays into two triplet excitons was a key source of these free triplets. \cite{Singh65}

However, although the possibility of singlet exciton fission  was recognized early on,
the fundamental physical mechanism for singlet-triplet conversion has  only recently attracted attention \cite{Zimmerman10,Zimmerman11,Chan11,Chan12,Chan13,Berkelbach13a,Berkelbach13b,Berkelbach14,Scholes15,Monahan15},
also prompted by the interest in its potential application  to solar energy harvesting \cite{Greyson10,Greyson10b,Monahan15}. 

Singlet exciton fission  can be described as \cite{Frankevich78, Pensack16, Breen17}
\begin{equation}
{}^1S_0 + {}^1S_1 \rightarrow {}^1(T_1 T_1) \rightarrow {}^1(T_1 \cdot \cdot \cdot T_1) \rightarrow {}^3T_1 + {}^3T_1   .\label{fissionreaction}
\end{equation}
Here the initial singlet excited state, represented  by a ground-state and excited-state molecule pair, first becomes a coherent geminate pair of triplets in an overall singlet state, ${}^1(T_1 T_1)$, which can then undergo a transition to a state ${}^1(T_1 \cdot \cdot \cdot T_1)$ with a larger distance between its triplet components \cite{Frankevich78,Breen17}, which then finally decays into two independent triplet excitons \cite{Scholes15}. Both intermediate triplet-pair states  can in general (depending on relative singlet and triplet energies) undergo geminate recombination, bringing the system back to a singlet state and allowing for photon emission. The final result of the fission process are two triplet excitons that independently diffuse in the crystal, and that, upon interacting with another triplet exciton, can undergo \emph{non-geminate} recombination, realizing the fusion process. In this work, we study the latter part of the fission process in rubrene single crystals,  providing a direct optical observation of an intermediate, coherent  multiexciton state. 

Quantum beats indicative of a coherent multiexciton state have been previously observed  in the nanosecond-scale PL emitted after pulsed excitation in tetracene molecular crystals.  \cite{Chabr81,Funfschilling85,Funfschilling85b,Burdett12,Burdett13,Wang15} Unfortunately, despite the current interest in  exciton fission  and the need to validate theoretical models, this  remained, to date, the only direct optical proof of such a multiexciton state.

\begin{figure}[t]
\includegraphics[width=8.5cm]{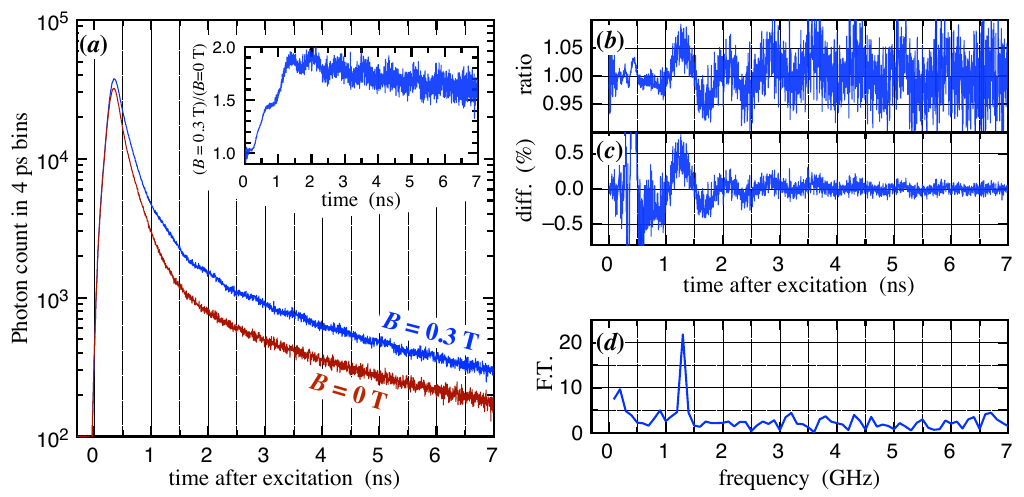}
\caption{Photoluminescence transients in a rubrene single crystal after impulsive photoexcitation. Excitation polarization along the $b$-axis of the crystal. Panel (a) shows the PL dynamics with and without an applied magnetic field of 0.3 T along the $c$-axis of the crystal, with the inset showing the ratio between the two measurements. Panels ($b$) and ($c$) show the PL after division by and subtraction of the non-oscillatory background. Panel ($d$) is a Fourier transform of the data in Panel ($b$).}
\label{beatsfig}
\end{figure}

We now report  PL quantum beats in rubrene single crystals (Fig.~\ref{beatsfig}) on the nanosecond time scale, which hints at a long-lived coherent multiexciton state that can undergo geminate recombination for several nanoseconds before it ultimately decays into independent triplets. This provides  important context for the recent work of Breen et al. \cite{Breen17}, who described a transition from  ${}^1(T_1 T_1)$ to  ${}^1(T_1 \cdot \cdot \cdot T_1)$ in less than 100 ps.

In rubrene,  triplet states have been observed to appear  on the picosecond time-scale after  photoexcitation \cite{Furube11, Ma12, Jankus13,Breen17}, and singlet fission is a dominant process that  leads to long-lived independent triplet excitons \cite{Ryasnyanskiy11}. In addition, the energy of the triplet states in rubrene is close to half that of the singlet states \cite{Herkstroeter81,Tsunoyama17}. This enables not only an efficient fission process, but also a  fusion process that leads to large enhancements of the steady state PL even under moderate excitation densities \cite{Biaggio13b,Irkhin16} and  can also be observed in triplet diffusion experiments \cite{Irkhin11}.

Rubrene single crystals were grown from 99\% pure rubrene powder from ACROS organics via standard vapor transport \cite{Laudise98} in a 1 inch diameter fused silica tube with a sublimation temperature of 300~$^\circ$C and an argon flow rate of 40 ml/min.  Following the established charge-transport 
literature \cite{Podzorov03,Sundar04} and earlier work \cite{Irkhin12}, we refer to the  two-fold rotation axis of the rubrene molecules in the crystal \cite{Jurchescu06} as the $c$-axis, with the $b$-axis corresponding to the crystal's molecular stacking direction, in which high charge-carrier  \cite{Podzorov03, Sundar04} and triplet exciton \cite{Irkhin11} mobilities have been observed. For this investigation, we selected homogeneous,  single crystal platelets with thicknesses of the order of $100\mu$m and mm-wide surfaces perpendicular to the $c$-axis. PL spectroscopy confirmed emission spectra characteristic of pristine rubrene, as given in Ref. \onlinecite{Irkhin12}, without the low energy PL emission band near $650$ nm that has sometimes been seen in other reports (see, $e.g.$, Refs.~ \onlinecite{Ma12,Ma13}).
For photoexcitation, we used $150$~fs laser pulses of $513$~nm wavelength and with a repetition rate of $200$~kHz,  obtained from a Light Conversion PHAROS laser. This excitation light was polarized along the $b$-axis of the crystals. We generally used a beam waist at the sample of $\sim40~{\mu}$m and peak fluences of up to $5 \times 10^{-2}$~Jm$^{-2}$, corresponding (for the 513 nm, $b$-polarized absorption length of $2.6 \ \mu$m \cite{Irkhin12}) to peak excitation densities of up to $\sim5 \times 10^{22}$ m$^{-3}$. Each sample was imaged during the measurements to confirm a high quality surface at the illumination spot. The PL was collected in a confocal setup and filtered by 550 nm long-pass filters that blocked the laser light but let most of the PL spectrum through \cite{Irkhin12}. The time dynamics was obtained using a time-correlated single photon counting system  from PicoQuant (Picoharp 300) using  integration times  on the order of one hour.
Neodymium permanent magnets were used to apply magnetic fields of up to 0.3 T, determined at the position of the sample to within $\pm 10\%$ using a Hall probe.

Fig.~\ref{beatsfig} summarizes our results for the PL dynamics obtained over three hours of integration time. The data measured in this experiment consists of the number of photons detected over time-intervals of 4 ps at various times after photoexcitation. Data was taken both with and without an applied magnetic field  of  0.3 T parallel to the $c$-axis; care was taken to leave other experimental parameters ($e.g.$ pulse fluence and illuminated spot) unchanged between measurements. The PL data generally shows a fast transient lasting a few hundred picoseconds, determined by the response time of our apparatus, followed by a slower decay on the nanosecond time-scale. The general PL dynamics on this time scale is not dissimilar to what was reported in Refs.~\onlinecite{Ma12} or \onlinecite{Piland13}. But our data  shows PL oscillations that are already visible in Fig.~\ref{beatsfig}$a$ and can be made evident by taking the ratio between the data with and without magnetic field (inset in Fig.~\ref{beatsfig}$a$), or by  plotting either the difference (Fig.~\ref{beatsfig}$b$) or the ratio (Fig.~\ref{beatsfig}$c$)  between the data and its non-oscillating background. The latter was obtained by fitting the data with a model function consisting of two exponential decays multiplied by a functional representation of the initial transient dominated by the instrument response function.

We assign the observed oscillations in the transient PL to quantum beats from a multiexciton state. These quantum beats are similar to those observed in tetracene \cite{Chabr81,Funfschilling85,Funfschilling85b,Burdett12,Burdett13,Wang15}, with the difference that we could not detect quantum beats in the absence of an applied magnetic field. We also note a transient magnetic-field induced enhancement of the background PL  (inset in Fig.~\ref{beatsfig}$a$), similar to that reported 
in Ref.~\onlinecite{Piland13}. In the following, however, we concentrate on the characterization of the  quantum beats.

\begin{figure}[t]
\includegraphics[width=8.5cm]{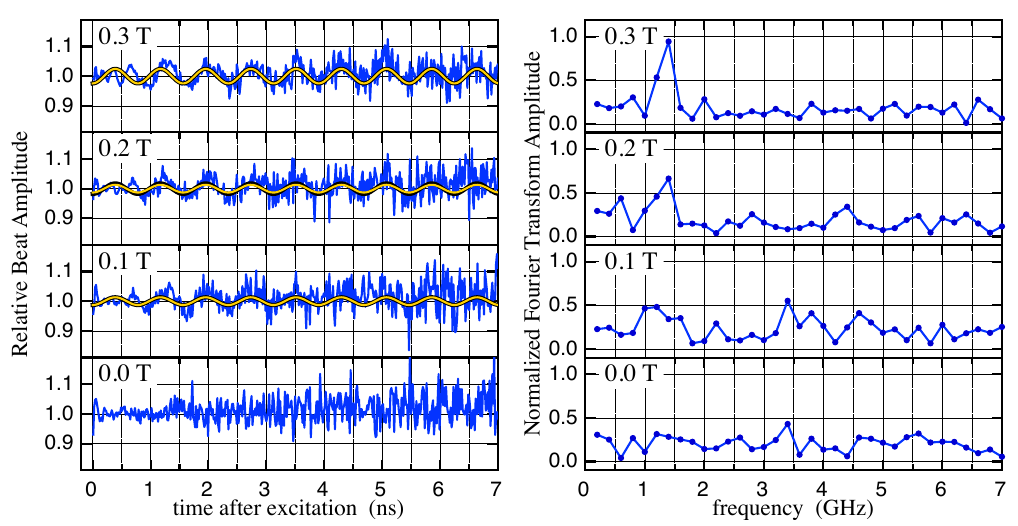}
\caption{Dependence of the extracted quantum beats on magnetic field strength.
Left: PL decay data divided by its non-oscillatory background, as obtained for different strengths of a magnetic field parallel to the $c$-axis of the crystal. The solid lines are plots of $[1-a \cos(2 \pi f t)]$ with $f=1.28$ GHz, obtained via a two-parameter fit to the data between 2 and 7 ns. The best-fit frequency remains $1.28 \pm 0.03$ GHz for all magnetic field strengths.
Right: Frequency density spectra as obtained by Fourier transforms of the data between 2 and 7 ns. }
\label{strengthdep}
\end{figure}

\begin{figure}[t]
\includegraphics[width=8.5cm]{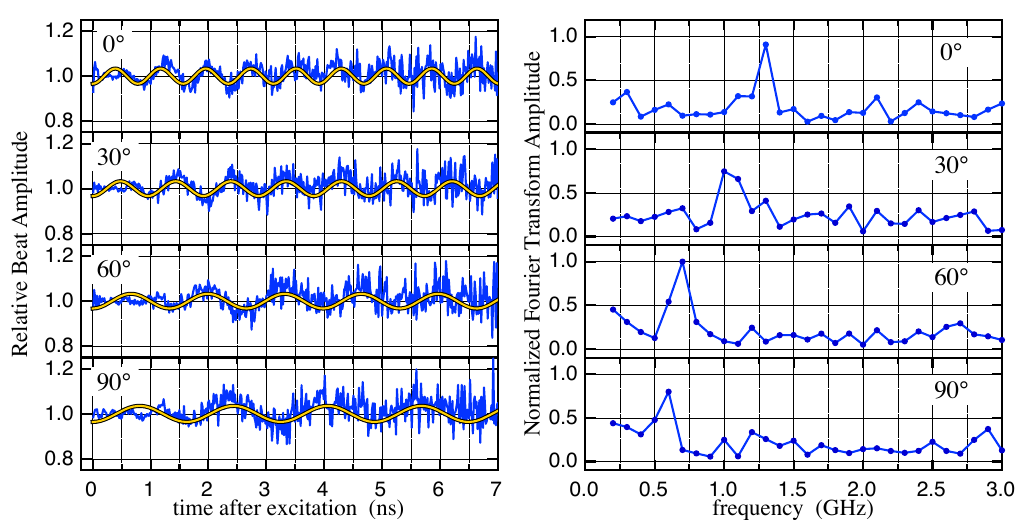}
\caption{Dependence of the extracted quantum beats on magnetic field orientation in the $b$-$c$-plane of the crystal.
Left: PL decay data divided by its non-oscillatory background. The angle between magnetic field and $c$-axis changes from $0^\circ$ to $90^\circ$ from the top trace to the bottom trace in steps of $30^\circ$. The solid lines are a two-parameter fit of a function of the form $[1-a \cos(2 \pi f t)]$ to the data between 2 and 7 ns.
Right: Frequency density spectra as obtained by Fourier transforms of the data between 2 and 12 ns. The frequencies $f$ as identified by  the functional fits  are $1.28$, $1.05$, $0.75$, and $0.62$ GHz, with an accuracy of $\pm 3\%$.
}
\label{angledep}
\end{figure}

Figs.~\ref{strengthdep} and ~\ref{angledep} present the dependence of the  quantum beats from the magnitude and orientation of the applied magnetic field. For clarity, the plotted data was smoothed using a running average over a 16 ps time interval. Both figures show the extracted quantum beats after an integration time of one hour, as obtained by dividing the data by the non-oscillatory background. We note that this method of extracting the quantum beats is  error-prone in the time-region where the PL changes very rapidly, dominated by the instrument response function. This creates artifacts in the data of Figs.~\ref{beatsfig}$b$, \ref{beatsfig}$c$, \ref{strengthdep} and \ref{angledep} below 1-2 ns. Because of this, we relied only on the data after 2 ns for all our analysis. To make the plots in Figs.~\ref{strengthdep} and \ref{angledep} clearer, we fitted the data in the 2-7 ns time interval using a simple sinusoidal function $[1-a \cos(2 \pi f t)]$ and plotted this function on top of the data.  From Fig.~\ref{strengthdep}, one can see that no quantum beats can be seen without an applied magnetic field, and that the quantum beats start being visible with an amplitude of a few percent of the non-oscillating background as the magnetic field strength increases past 0.1 T. For all these measurements, the frequency of the beats remains the same within the experimental error, pinned at $1.28 \pm 0.03$ GHz based on the above mentioned cosine fits to the data.

The orientation of the magnetic field, on the other hand, does affect the frequency of the beats. This is shown in Fig.~\ref{angledep}, where the same type of data as Fig.~\ref{strengthdep} is plotted for varying magnetic field directions instead of magnetic field magnitude (kept at $0.24 \pm 0.02$ T for this measurement). The quantum beats are dominated by a single frequency that decreases from $1.3$ GHz for a magnetic field along the $c$-axis of the crystal to $0.6$ GHz for a magnetic field parallel to the $b$-axis, as seen both from the sinuosoidal fits to the data and the position of the peak in each Fourier transform. 
We note that the beats at all frequencies can be well fitted by the function $[1-a \cos(2 \pi f t)]$, which is constrained to have a minimum at time zero,  set for all experiments to be within 0.1 ns from the time of arrival of the excitation pulse. Allowing for an additional phase or additional frequency terms only leads to overfitting, with no significant improvement in fitting quality.

The long lifetime of independent triplet excitons in rubrene \cite{Ryasnyanskiy11} implies that the singlet states created by each of our excitation pulses, and therefore the multiexciton state, exist in the presence of a  high density of triplet excitons established by accumulating the contributions from the previous excitation pulses.
Because of efficient fission, at our highest excitation density each pulse adds a triplet density of the order of twice the excitation density, $\sim10^{23}$ m$^{-3}$ . Any decay in triplet density over 5 microseconds (the time interval between our pulses) is due to triplet-triplet recombination, and and at these densities is expected to be small over the time-interval between excitation pulses \cite{Ryasnyanskiy11,Ward15}. Under our experimental conditions the PL power detected 20 ns after excitation (when the density of multiexciton states able to undergo geminate recombination has decayed to less than 10\% of its initial density and the fission into independent triplets can be regarded as complete), was only $\sim3$ times larger than that measured before the arrival of the next pulse ($\sim5 \ \mu$s after excitation). Since at these times the PL is dominated by triplet fusion and proportional to the square of the triplet density \cite{Ryasnyanskiy11,Ward15}, this data means that the triplet density found 20 ns after an excitation pulse is about $\sim1.5 \approx \sqrt{3}$ larger than the triplet density already found in the crystal before photoexcitation, which then must be of the order of twice the triplet density generated by each pulse, $\sim2 \times 10^{23}$~ m$^{-3}$. This corresponds to an average distance between triplet excitons of the order of $\sim20$ nm. This estimation is consistent with what would be expected under continuous wave illumination of similar average intensity \cite{Biaggio13b}.
The triplet diffusion constant of $D \sim1.6 \times 10^{-7}$ m$^2$s$^{-1}$ in rubrene \cite{Irkhin16} then means that diffusing triplets could interact with each other and the multiexciton state within a  time $(20 {\rm \ nm})^2/D$, or a few nanoseconds. This implies the possibility of an interaction between background triplet population and multiexciton state, and also an interaction between the new triplet states created by the excitation pulse, with the related non-geminate recombination contributing to the PL dynamics PL dynamics later on the nanosecond scale. It may be  that no quantum beats have been observed in rubrene in Ref.~\onlinecite{Ma12} because they were using a repetition rate of 80 MHz at higher excitation densities, and in Ref.~\onlinecite{Piland13} because of signal-to-noise issues in an experiment performed at a 40 kHz repetition rate with lower excitation densities than used here. But we should also mention that while all the data we present here has been taken at 200 kHz, we  could obtain essentially the same quantum beats after reducing the repetition rate by a factor of four and increasing the  integration time by the same amount.

\begin{figure}[t]
\includegraphics[width=8.5cm]{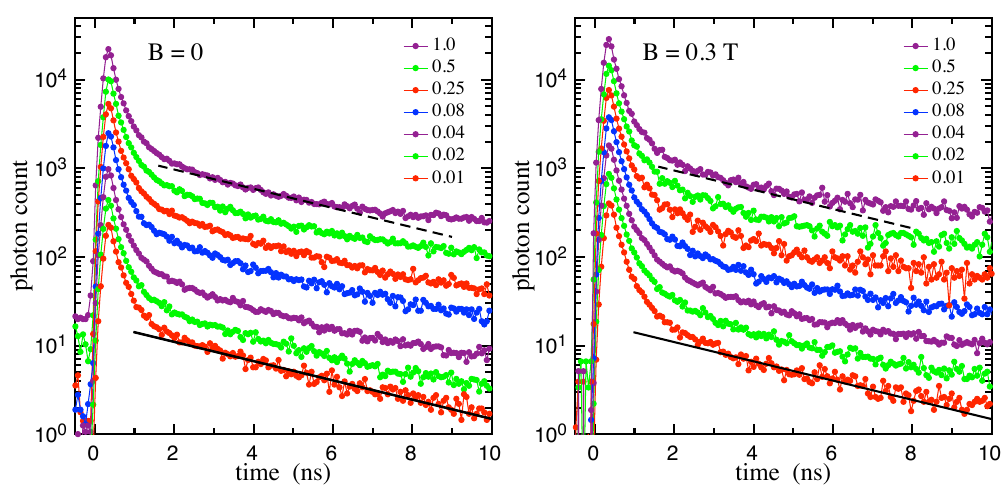}
\caption{
Variation in the PL transients with excitation density in two different samples, without and with an applied magnetic field of 0.3 T. The excitation densities are given in the legends in terms of ratios to the highest excitation density used, which was of the order of  $\sim5 \times 10^{23}$ m$^{-3}$. As the contribution of photoluminescence originating from triplet-triplet interaction decreases with excitation density, a clearer exponential decay emerges. The two solid lines represent exponential decays with a time constant of 4 ns that fit the lowest excitation density. The dashed lines are exponential decays with a 4ns time constant that highlight the deformation of the PL decay observed at higher excitation densities.  No quantum beats are observed at the lowest excitation densities because of the limited integration time used here.}
\label{fournsdata}
\end{figure}

The next question to address is how long the PL from geminate recombination lasts, and if the quantum beats change during this time. From  Figs.~\ref{strengthdep}-\ref{angledep} and previous work \cite{Ryasnyanskiy11}, the PL decay is  characterized by a fast transient followed by a slower decay that tends to a power-law, caused by non-geminate recombination of independent triplets. The quantum beats occur during the initial fast transient and their relative amplitude appears to remain constant until they vanish into the increasing noise due  the decreasing signal levels. To investigate the lifetime of the multiexciton state, it is necessary to make sure that non-geminate recombination of independent triplets does not affect the measurements at least during the first $10$~ns. We did this by measuring PL transients while lowering the excitation density. The results are shown in Fig.~\ref{fournsdata}. The decays measured at higher excitation densities are strongly non-exponential, with a flattening out of the decay on the semilogarithmic plots as non-geminate recombination starts to contribute to the detected PL. But  the decays become more exponential at lower excitation densities as non-geminate recombination becomes less probable on this time-scale. This can be seen both in the data measured without applied magnetic field, and in that measured with applied field. The exponential time-constant of the single exponential transient that emerges at low excitation densities in Fig.~\ref{fournsdata} is $4.0 \pm 0.2$ ns, consistent with an earlier observation \cite{biaggio13}. This can be seen as the effective lifetime of a multiexciton state that decays by dissociating into two independent triplet excitons, as required both by the onset of the power-law PL decay at later times in the higher-excitation PL transients in Fig.~\ref{fournsdata}, and the demonstrated high efficiency of singlet fission \cite{Ryasnyanskiy11,Biaggio13b}. A constant relative amplitude of the quantum beats during this decay (See Figs.~\ref{beatsfig}-\ref{angledep})  would be consistent with an expected absence of dephasing on this time-scale. In any case, the 4 ns exponential decay that we have identified can be regarded as another signature of the multiexciton state and should be the focus of further experiments.

What is now the nature of this multiexciton state? The simplest  explanation is that it is one of the intermediate states in Eq.~\ref{fissionreaction}. Our data cannot address any possible electronic coherence between the photoexcited singlet state and the  triplet-pair state \cite{Tamura15}, but the quantum beats prove both the coherence of the multiexciton state as well as its  capability to undergo  geminate recombination. It is possible that this multiexciton state is the ${}^1(T_1 \cdot \cdot \cdot T_1)$  state arising from an initial spatial separation of the triplets on the picosecond time scale \cite{Frankevich78,Breen17}, but the  description in  Ref.~\onlinecite{Breen17} of  this transition  as an ``irreversible separation process'' must not be interpreted as forbidding geminate recombination.  Rather, we have shown that the probability of geminate recombination decays over several nanoseconds. The probability for geminate recombination  should depend on the spatial as well as on the spin wavefunction, but we cannot tell what role, if any, random spin flips might play on this time scale. Still, we have shown above that the multiexciton state decays by dissociating into separate triplets, and our data would be consistent with a  ${}^1(T_1 \cdot \cdot \cdot T_1)$ bound-state, characterized by a constant average distance between the triplet states, that has a constant probability of undergoing a transition to effectively separated, independent triplet excitons. Alternatively, it would also be conceivable  that the triplet separation in each ${}^1(T_1 \cdot \cdot \cdot T_1)$ state continues increasing on the nanosecond time scale until geminate recombination becomes irrelevant and interaction with other triplets dominates.

Another observation is that we did not see quantum beats in the absence of a magnetic field and that we only detected single-frequency beats in a magnetic field. It is possible that the beat amplitude without magnetic field was just too small to  emerge from noise: a significant difference between beat amplitudes  with and without magnetic fields has  been previously observed in tetracene \cite{Chabr81}. The fact that quantum beats under an applied magnetic field have a single frequency is expected. In the simplest model, the  triplet-pair in an overall singlet state is proportional to $|1-1\rangle+|-1\ 1\rangle + |00\rangle$, where $|m_1 m_2\rangle$ indicates the pair-state where the spin components along the magnetic field are $m_1$ and $m_2$. The energy difference between the $|1-1\rangle+|-1\ 1\rangle$ state and the $|00\rangle$ state \cite{Swenberg73}  then leads to a single frequency oscillation of the probability for geminate recombination  \cite{Chabr81,Burdett13}. To estimate this frequency, we can assume that the triplet energies of a rubrene molecule are determined by the tetracene backbone, and use the zero-field splitting parameters $D = 0.052$ cm$^{-1}$ and $E = - 0.0052$  cm$^{-1}$ of the tetracene molecule \cite{Yarmus72} to find the energy difference. Given rubrene's crystal structure, a magnetic field along the $c$-axis is  parallel the two-fold rotation axis of every rubrene molecule in the crystal, which corresponds to the the short axis, ``y'', of the tetracene backbone. In this case the energy gap can be calculated \cite{Swenberg73} to be    $D + 3 E$, equal to a frequency of $1.1$ GHz, not too far from the frequency observed in Figs.~\ref{beatsfig}-\ref{angledep} for $\vec B \parallel c$. But this energy difference changes for other magnetic field orientations, which causes the angle-dependence of the beat frequency seen in Fig.~\ref{angledep}.

Combining all of the above, we  explain our experiments with a coherent multiexciton state that is generated by singlet fission and has an effective lifetime of 4 ns. This state may correspond to the recent observation of Ref.~\onlinecite{Breen17}, or in general to the ``diffusive correlated triplet-pair'' hypothesized early on in Ref.~\onlinecite{Frankevich78}. But the exact mechanism with which this state dissociates into triplet excitons that will then diffuse independently and separately in the crystal is still an open question.

Finally, we note that  the amplitude of the quantum beats we detected is a few percent of the non-oscillatory background. There is currently not enough data to assess this amplitude or to discuss why it is lower than in tetracene. The beat amplitude depends on the transition probability amplitudes between the sublevels of the triplet-pair state as they evolve in time and the singlet state, and is expected to depend on both magnitude and orientation of the magnetic field \cite{Wang15}, but in this work we could only explore a limited range of magnetic field strength and directions. In addition, beats visibility is affected by phonons \cite{Chan13,Monahan15,Bakulin16}, and it could also be affected by the background of independent triplets mentioned earlier, which is expected to be stronger in rubrene than in tetracene because triplets in rubrene have a longer  lifetime \cite{Ryasnyanskiy11}.

In conclusion, we observed periodic modulations in the photoluminescence dynamics of impulsively excited rubrene single crystals that can be attributed to quantum beats originating from a  multiexciton state. We established that this state consists of a spin-coherent triplet-pair and that it acts as an intermediate state towards fission into separate, independent triplet excitons. We also identified another signature of this multiexciton state that can be observed at lower excitation densities:  a $4$ ns exponential decay in the PL background accompanying the quantum beats.

\bibliography{RubreneEtAl} 
\end{document}